# Variation of polarization distribution of reflected beam caused by spin separation

Yu Jin, Zefang Wang, Yang Lv, Hao Liu, Ruifeng Liu, Pei Zhang, Hongrong Li, Hong Gao,[*] and Fuli Li

*Department of Applied Physics, Xi'an Jiaotong University, Xi'an 710049, China*
[*]*honggao@mail.xjtu.edu.cn*

**Abstract:** The variation of polarization distribution of reflected beam at specular interface and far field caused by spin separation has been studied. Due to the diffraction effect, we find a distinct difference of light polarization at the two regions. The variation of polarization distribution of reflected light provides a new method to measure the spin separation displacement caused by Spin Hall Effect of light.

## 1. Introduction

The Spin Hall Effect of light (SHEL) at the interface has attracted much attention for its potential application in quantum information and optical information processing [1-5]. SHEL was first proposed in 2004, and it describes the opposite transverse displacements of two spin components of photonics at the interface [1]. The SHEL effect can be seen as the consequence of Berry phase, which corresponds to the spin-orbit interaction. Weak measurement has been applied to measure the displacement [4, 5]. Recently, the in-plane spin separation (IPSSL) has also been reported, which can be seen as essential composition of the theory of spin separation [6]. Polarization variation of light at the interface caused by SHEL has some crucial features and has been mentioned in Refs.[3, 6]. Previous studies mainly focus on the polarization distribution at the interface, however the polarization property between the interface and far field are intrinsically different due to diffraction effect of two separated spin polarizations. The effect of beam propagation is implied in the formalism developed in [7], however this has not been made explicit.

In this paper, we have systematically studied the polarization property of the reflected beam and demonstrated the variation of polarization distribution from the interface to far field. The polarization distribution at interface is determined by spin separation (SHEL and IPSSL), while the polarization variation during the propagation is caused by diffraction effect. In our theory, upon the reflection at an air-glass interface, a linearly polarized incident light becomes a beam with elliptically polarized distribution. However, during the propagation after reflected, the polarization ellipticity angle gradually vanish and the major axis rotate. At far field, the reflected beam eventually evolves to linearly polarized distribution. This feature provides a feasible way to measure the displacement caused by SHEL. Our experimental results show a perfect agreement with the theoretical prediction.

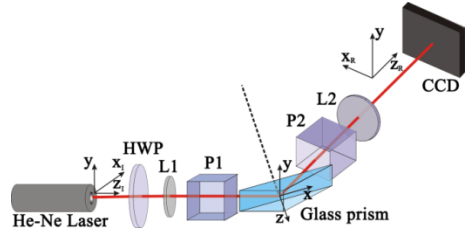

Fig. 1. Experimental setup and coordinate systems for studying the polarization distribution. The He-Ne laser generates a fundamental Gaussian beam with the wavelength of 632.8nm. HWP, half-wave plate for adjusting the intensity after P1. L1 and L2, lenses with 100mm and 150mm focal lengths, respectively. P1 and P2, Glan polarizers. In our experiment we set the reflect interface at the back focal plane of L1 and at the front focal plane of L2. Our observation plane (CCD) is chosen at the back focal plane of L2. The refractive index of glass prism is 1.49.

## 2. Theoretical analysis

### 2.1 General description of the reflected beam

To ensure the integrity of our theory, it is necessary to review the general description of spin separation at interface [4, 6]. The general SHEL experimental setup and coordinate systems are shown in Fig.1. We consider the wave-packets containing distribution of wave vector $\vec{k}^{(I)}$ and $\vec{k}^{(R)}$ for incident beam and reflected beam, which can be expressed as: $\vec{k}^{(I)} = k_{x_I}\hat{x}_I + k_y\hat{y}_I + k_I\hat{z}_I$ and $\vec{k}^{(R)} = k_{x_R}\hat{x}_R + k_y\hat{y}_R + k_R\hat{z}_R$ with $|k_{x_{I,R}}|, |k_{y_{I,R}}| = k_{I,R}$. $\hat{x}_{I,R}$, $\hat{y}_{I,R}$ and $\hat{z}_{I,R}$ are the unit vectors along $x_{I,R}$, $y_{I,R}$ and $z_{I,R}$ axis, and $k_I = 2\pi/\lambda$ with $\lambda$ is the wavelength of light in the air. Since $\hat{y}_{I,R}$ are always vertical direction, they can replaced by $\hat{y}$ thereafter. We can get $k_I = k_R$ from Snell's law. For arbitrary incident wave packet, we have $k_{x_I} = -k_{x_R}$ and $k_{y_I} = k_{y_R}$. Considering that the incident beam is not exact planar wave, we can express Fresnel reflection coefficients, $r_s$ and $r_p$, as: $r_s = r_{s\theta_I} + (k_{x_I}/k_I)\partial r_s/\partial \theta_I$, $r_p = r_{p\theta_I} + (k_{x_I}/k_I)\partial r_p/\partial \theta_I$, where $\theta_I$ is incident angle, $r_{s\theta_I}$ and $r_{p\theta_I}$ are s- and p-wave reflection coefficients of planar wave at incident angle of $\theta_I$.

Under the action of the interface, horizontally polarized incident beam $|H\rangle$ and vertically polarized incident beam $|V\rangle$ evolve as:

$$\left|H(\vec{k}^{(I)})\right\rangle \to r_{p\theta_I}(\left|H(\vec{k}^{(R)})\right\rangle + k_{x_R}\Delta^H\left|H(\vec{k}^{(R)})\right\rangle + k_y\delta^H\left|V(\vec{k}^{(R)})\right\rangle \quad (1.a)$$

$$\left|V(\vec{k}^{(I)})\right\rangle \to r_{s\theta_I}(\left|V(\vec{k}^{(R)})\right\rangle + k_{x_R}\Delta^V\left|V(\vec{k}^{(R)})\right\rangle - k_y\delta^V\left|H(\vec{k}^{(R)})\right\rangle \quad (1.b)$$

where

$$\delta^H = \cot\theta_I(-1 - r_{s\theta_I}/r_{p\theta_I})/k_I, \quad \delta^V = \cot\theta_I(-1 - r_{p\theta_I}/r_{s\theta_I})/k_I,$$

$$\Delta^H = -1(r_p/r_{p\theta_I} - 1)/k_{x_I}, \quad \Delta^V = -1(r_s/r_{s\theta_I} - 1)/k_{x_I}.$$

$\delta^H$ and $\delta^V$ account for SHEL, while $\Delta^H$ and $\Delta^V$ account for in-plane spin separation. When a horizontally polarized light with a certain intensity distribution incident on the interface, from Eq.(1.a), the reflected beam then can be expressed as:

$$\left|\phi_{final}(k_{x_R}, k_y)\right\rangle = r_{p\theta_I}\phi_{initial}(k_{x_R}, k_y)(\left|H\right\rangle + k_{x_R}\Delta^H\left|H\right\rangle + k_y\delta^H\left|V\right\rangle), \quad (2)$$

where, $\phi_{initial}(k_x, k_y)$ and $\phi_{final}(k_x, k_y)$ are the wave functions of incident and reflected beam at interface in momentum space, respectively. The above equation can also be expressed in spin basis to get more physical meaning as described in Refs.[4, 6]. Then Fourier transformation gives the spatial distribution of reflected beam at the interface as:

$$\left|\psi_{final}(x,y)\right\rangle = \frac{r_{p\theta_I}}{\sqrt{2}}\psi_{initial}(x - i\Delta^H, y - \delta^H)\left|+\right\rangle + \frac{r_{p\theta_I}}{\sqrt{2}}\psi_{initial}(x - i\Delta^H, y + \delta^H)\left|-\right\rangle$$

$$= \frac{r_{p\theta_I}}{2}[\psi_{initial}(x - i\Delta^H, y - \delta^H) + \psi_{initial}(x - i\Delta^H, y + \delta^H)]\left|H\right\rangle \quad (3)$$

$$+ \frac{r_{p\theta_I}i}{2}[\psi_{initial}(x - i\Delta^H, y - \delta^H) - \psi_{initial}(x - i\Delta^H, y + \delta^H)]\left|V\right\rangle,$$

where, $\psi_{initial}(x,y)$ and $\psi_{final}(x,y)$ are the wave functions of incident and reflected beam at spatial space, respectively. $\left|H\right\rangle = \frac{1}{\sqrt{2}}(\left|+\right\rangle + i\left|-\right\rangle)$ and $\left|V\right\rangle = \frac{1}{\sqrt{2}}(\left|+\right\rangle - i\left|-\right\rangle)$. To obtain Eq.(3) we assume $k_{x_R}\Delta^{H(V)}, k_y\delta^{H(V)} = 1$. Following the same procedure, the description of reflected beam with vertical polarization can also be obtained, and the result is similar.

*2.2 Variation of polarization distribution of the reflected beam*

In this section, we only focus on the horizontally polarized incident beam and discuss its polarization properties. The result of vertically polarized is quite similar. Arbitrary polarization beam can be obtained by the linear combination of horizontal and vertical polarization. For realistic, the incident beam has a Gaussian shape as $\psi_{initial}(x,y) \propto e^{-\frac{x^2+y^2}{w_0^2}}$ or $\phi_{initial}(k_x, k_y) \propto e^{-\frac{w_0^2(k_x^2+k_y^2)}{4}}$. To better describe the polarization properties of the reflected beam, we use two parameters, the orientation angle $\gamma$ and the ellipticity angle $\chi$. $\gamma$ denotes the angle between the major semi-axis of the ellipse and the horizontal (x-axis) direction. $\chi = arc\cot(\varepsilon)$, where $\varepsilon$ is the ellipticity defined as the major-to-minor-axis ratio. Therefore $\gamma$ and $\chi$ can be calculated by the following equations:

$$\tan(2\gamma) = \frac{2\text{Re}(E_H E_V^*)}{|E_H|^2 - |E_V|^2}, \quad \tan(2\chi) = \frac{2\text{Im}(E_H E_V^*)}{\sqrt{(|E_H|^2 - |E_V|^2)^2 + (2\text{Re}(E_H E_V^*))^2}}. \quad (4)$$

$E_H$ and $E_V$ are the horizontal and vertical components of the reflected beam's wave function, respectively. Using Eqs.(3) and (4), the polarization properties of reflected beam at interface can be obtained. Figure 2(a) and (c) show the polarization distribution of reflected light at a certain position of the beam cross section, where (a) is the schematic plot and (c) is the calculation results. When $y=0$, it is horizontally polarized. At the other position, it is elliptically polarized and the major-axis is parallel to $x_R$-axis. Their spin directions are different between the two sides of $y=0$.

After the reflect beam propagates for a long distance $z_R$ ($k_I w_0^2/(2z_R) = 2\pi$), by using Fraunhofer diffraction theory, we can describe its property as follows:

$$\psi(x_R, y, z_R) = \frac{e^{ik_R z_R}}{i\lambda z_R} e^{ik_R \frac{x_R^2+y^2}{2z_R}} \mathrm{F}\{\psi_{final}(x,y)\} = \frac{e^{ik_I z_R}}{i\lambda z_R} e^{ik_I \frac{x_R^2+y^2}{2z_R}} \phi_{final}(k_I \frac{x_R}{z_R}, k_I \frac{y}{z_R}). \quad (5)$$

F denotes Fourier transformation. Using Eqs.(2) and (5), the polarization properties of reflected beam at far field can be revealed and they are shown in Fig.2(b) and (d). Unlike the results at the interface, the polarization distribution of reflected beam becomes pure linear at a certain position of the beam cross section. The polarization at $y=0$ is still horizontally polarized, while the polarization at other position has rotated a small angle. And the directions of rotation are different between two sides of $y=0$.

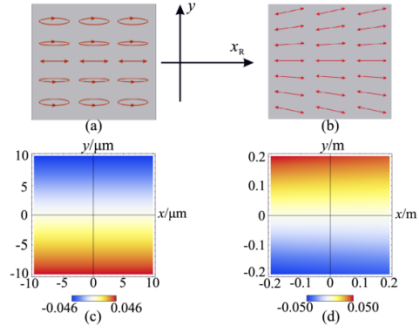

Fig. 2. The results of theoretical calculation of the polarization properties at interface (a), (c) and at far field (b), (d). (a) and (b) are schematics to show the polarization property. (c) shows the calculation value of $\chi$ (rad) at the interface and (d) shows the calculation value of $\gamma$ (rad) after 10m propagation for the reflected beam. Note that the incident beam is horizontally polarized and beam waist at the interface is 10 μm.

To show how the polarization of light evolves from interface to far field, we use Fresnel diffraction to describe the propagation of reflected beam. This gives: $\phi(k_{x_R}, k_y, z_R) = \phi_{final}(k_{x_R}, k_y) e^{ik_I z_R} e^{-iz_R(k_x^2+k_y^2)/(2k_I)}$. According to Eq.(2) and conducting an inverse Fourier Transformation, we can get the wave function of the reflected beam after propagating of distance $z_R$:

$$\psi(x_R, y, z_R) \propto r_{p_{\theta_I}} \frac{k_I w_0^2 e^{-\frac{k_I(x_R^2+y^2)}{k_I w_0^2 + 2iz_R}}}{(k_I w_0^2 + 2iz_R)^2} \{[k_I(w_0^2 - 2i\Delta^H x_R) + 2iz_R]|H\rangle - 2i\delta^H k_I y|V\rangle\}, \quad (6)$$

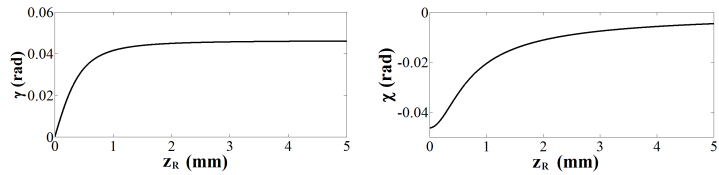

Fig. 3. Variation of polarization distribution during the propagation with horizontally polarized incident beam. Since the trend is similar, the figure only shows the polarization property at a certain position: $x_R=0$, $y$ =radius of beam waist at distance $z_R$. In our calculation, $w_0=10$μm and incident angle=45°.

where $w_0$ is beam waist at the interface. By using Eq.(6), the variation of polarization distribution can be revealed, and the results are shown in Fig.(3).The elliptically polarized beam quickly becomes quite linearly like polarized beam in a relatively short distance. This distance is related to dimension of beam waist. At far field, the light is purely linearly polarized. Note that at the different position of the beam cross section, their polarization direction is different, which can be seen from Fig. 2(b).

*2.3 Measurement of y-displacement and intensity distribution at the back focus plane*

To validate the polarization of reflected beam becomes purely linear at far field, we introduce a lens L2 with a focal length of *f*. We set the reflect interface at the front focal plane of L2 and the observation plane at its back focal plane. Considering the Fourier Transformation of L2, the wave function at its back focal plane can be expressed as: $\psi(x_R, y, z_f) \propto F\{\psi_{final}(x_R, y, 0)\} = \phi_{final}(k_I \frac{x_R}{f}, k_I \frac{y}{f})$, which is similar to Eq.(5). Thus the polarization is also linear at the back focal plane, which is equivalent to far-field when L2 hasn't been introduced. According to Eqs.(2) and (5), the polarization of the reflected beam at observation plane can be expressed by the polarization angle, $\theta$:

$$\tan\theta = k_I \frac{y}{f} \delta^H / (1 + k_I \frac{x_R}{f} \Delta^H). \tag{7}$$

The above equation shows the polarization angle at a certain position of observation plane is determined by the position and displacements. When the incident beam is horizontally polarized, the polarization angle at $x_R$-axis is horizontal. If we introduce a polarizer P2 with vertical polarizing direction, there will be a dark fringe at $x_R$-axis, as show in Fig.4(a) and (d).

For simplicity, we just consider the property at the *y*-axis. Thus the polarization angle satisfies $\tan\theta = k_I y \delta^H / f$, which indicates the polarization is related to *y* and the displacement ($\delta^H$) of SHEL. Meanwhile the dark fringe moves along the *y*-axis while rotating the P2. The rotating angle ($\Delta\theta$) and the distance of the dark fringe move ($\Delta y$) have a relation and it can be approximately expressed by (we choose $\Delta\theta = 0$ when the dark fringe is in the middle of the intensity profile):

$$\Delta\theta = k_I \frac{\Delta y}{f} \delta^H \quad \text{or} \quad \Delta y = \frac{\Delta\theta}{\delta^H k_I} f. \tag{8}$$

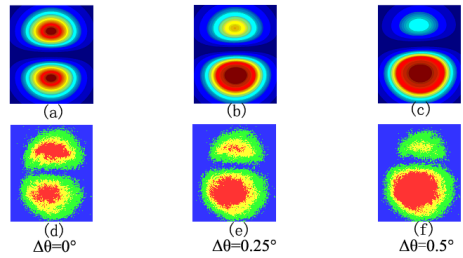

Fig. 4.Intensity profiles of reflected beam at the back focal plane of lens L2. (a),(b) and (c) show the theoretical results. (d),(e) and (f) are the experimental results.

Thus the spin separation displacement ($\delta^H$) of SHEL can be obtained by knowing $\Delta y$ and $\Delta\theta$. We can also use the polarization property to obtain the intensity distribution at the observation plane (back focal plane of L2). It can be expressed as:

$$I(x_R,y,f) \propto |e^{-w^2(k_I^2 x_R^2 + k_I^2 y^2)/(4f^2)}[\delta^H k_I y cos(\Delta\theta)/f + (1+\Delta^H k_I x_R/f)sin(\Delta\theta)]|^2. \quad (9)$$

Thus the intensity profiles of reflected beam at back focal plane with different rotating angle $\Delta\theta$ of P2 can be obtained, and the theoretical results are shown in Fig.4(a, b, c).

### 3. Experiment

From Theoretical analysis above, we know SHEL has great influence on the polarization distribution. According to Eq.(7) and Eq.(8), an amplified factor of $k_I y/f$ converts $\delta^H$ into variation of polarization distribution. Thus when rotating P2, we can notice considerable shift of dark fringe. Equation (8) provides a feasible way to measure the spin displacement of SHEL, which is based on the measurement of polarization distribution of observation plane. Then we accomplish an experiment to verify our theoretical prediction. The experimental setup is shown in Fig.1. The light with 632.8nm generated by He-Ne laser passes through HWP and P1, and becomes horizontally polarized. When the beam reflects from the interface, it suffers from polarization evolution as discussed above. We use a charge-coupled device (CCD) to record different intensity profiles when we rotate the P2. The results are shown in Fig.4(d, e, f). Clearly, the dark fringe moves as $\Delta\theta$ changes. We then measure the shift of dark fringe along the $y$-axis, $\Delta y$, by processing the pictures we get from CCD. The experimental data are showing in Fig.5. The solid lines are the theoretical calculation based on Eq.(8) by using our experimental parameters, where $\delta^H$ at different incident angles are calculated from Eq.(1). The experimental data and theory agree perfectly.

It should be mentioned that previous measurement of SHEL displacement requires position sensitive detector (PSD) to measure the gravity center of the reflected light. In our experiment, we measure the displacement by just using CCD to record pictures and analyze shift of dark fringe along the y-axis.

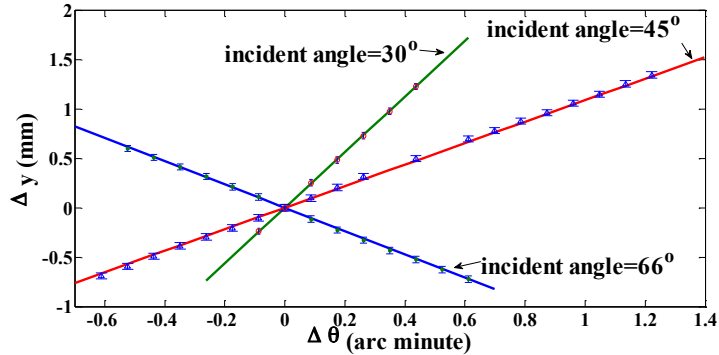

Fig. 5. Relation between $\Delta y$ (the distance dark fringe move along y-axis) and $\Delta\theta$ (the angle rotate away from vertical direction) for horizontally polarized incident beam. The circles, triangles and dots are experimental data at three incident angles: 30º, 45º and 66º. The solid lines are theoretical results. Here SHEL displacements are 89.5nm, 231.5nm and -213.3nm for the three incident angles 30º, 45º and 66º, respectively.

### 4. Conclusion

The variation of polarization distribution of reflected beam at the interface and far field caused by spin separation has been studied. We find a distinct difference of light polarization between the two regions due to the diffraction effect. The polarization evolution of light also provides a new method to measure the spin separation displacement caused by Spin Hall Effect of light. Our experimental results exhibit good agreement with the theoretical prediction.

**Acknowledge**

We thank Dr. Dong Wei for very useful discussion on this work. We acknowledge financial support from the National Natural Science Foundation of China (NSFC) under grants 11074198, 11004158, 11174233, and Special Prophase Project on the National Basic Research Program of China under grants 2011CB311807.